\documentclass{revtex4}
\usepackage{psfig}
\usepackage{subfigure}
\usepackage{graphicx}
\usepackage{doublespace}
\usepackage{amsmath}
\usepackage{amsfonts}

\begin{document}


\title{Synchronization of hypernetworks \\ of coupled dynamical systems}

\author{Francesco Sorrentino \\ 
Dipartmento per le Tecnologie, Universit{\`a} degli Studi di Napoli Parthenope, 80143 Napoli, Italy. \\ E-mail: {\tt fsorrent@unina.it}.}%

\begin{abstract}
We consider synchronization of coupled dynamical systems when different types of interactions are simultaneously present.
We assume that a set of dynamical systems are coupled through the connections of two or more distinct
networks (each of which corresponds to a distinct type of interaction), and we refer to such a system as a dynamical hypernetwork. 
{Applications include neural networks formed of both electrical gap junctions and chemical synapses, the coordinated motion of shoals of fishes communicating through both vision and flow sensing, and hypernetworks of coupled chaotic oscillators.}
 We first analyze the case of a hypernetwork formed of $m=2$ networks.   We look for necessary and sufficient conditions for synchronization. We attempt at reducing the linear stability problem in a master stability function form, i.e., at decoupling the effects of the coupling functions from the structure of the networks.
Unfortunately, we are unable to obtain a reduction in a master stability function form for the general case. However, we show that such a reduction is possible in three cases of interest:
(i) the Laplacian matrices associated with the two networks commute;
(ii) one of the two networks is unweighted and fully connected;
(iii) one of the two networks is such that the coupling strength from node $i$ to node $j$ is a function of $j$ but not of $i$. Furthermore, we define a class of networks such that if either one of the two coupling networks belongs to this class, the reduction can be obtained independently of the other network. 
As an example of interest, we study synchronization of a neural hypernetwork for which the connections can be either chemical synapses or electrical gap junctions. We propose a generalization of 
our stability results to the case of hypernetworks formed of $m\geq 2$ networks. 
\end{abstract}

\maketitle

pacs {05.45.Xt} {Synchronization, coupled oscillators}

pacs {05.45.Pq} {Chaotic systems}

pacs {89.75.-k} {Complex systems}

\section{Introduction}

Synchronization of coupled dynamical systems has been the subject of a considerable amount of research  (see e.g., \cite{FujiYama83,Pe:Ca,Ding:Ott,SUCNS,Report2}) with applications ranging from adaptive synchronization strategies \cite{So:Ott:Day,ADAP1,ADAP2,ADAP3,ADAP4,ADAP5} to pinning control \cite{Pinn1,Pinn2,PinnA,PinnIEEE}. One case of interest is that of complete synchronization that occurs when 
the individual systems, if appropriately coupled, converge on the same time-evolution. Complete synchronization can be observed in the presence of selective coupling, i.e., the systems are coupled through the connections of a network. A common underlying assumption is that the interactions among  the systems are all of the same type.
For this case, it has been shown that stability of the synchronized state depends on the details of the underlying network topology.

{In this framework, the master stability function (MSF) approach \cite{Pe:Ca} to synchronization of networks of coupled identical dynamical systems has been widely investigated in the literature \cite{Ba:Pe02,NSG,Pecora2009,SOPO}. An outstanding problem is how to obtain a reduction of the stability problem in a MSF form when the set of coupled dynamical systems simultaneously interact through different networks, with each network being associated with a  distinct coupling function.}

In this paper, we will focus on complete synchronization and we will retain selective coupling but 
we will allow for different types of couplings between the systems. We assume that all the connections that correspond to the same type of coupling form a network and the systems are connected by more than one network. This case is relevant to any situation where the individual units are allowed to interact through different types of coupling.  
{For example, neurons in the brain are  connected through both  electrical gap junctions and chemical synapses, see e.g., \cite{IZH,Ad:Pr:Dh}. The coordinated motion of shoals of fishes depends on the sensory capabilities of each individual fish.  Fishes typically use
vision but also chemical/flow sensing in order to localize their mates and coordinate their individual motion with respect to the shoal \cite{Pa:Pi,Ab:Po} (as in other animal species, the number of neighbors that can be simultaneously sensed by each fish is typically bounded and depends on the specific kind of interaction \cite{Krause}).
Another example is that of interdependent networks, such as e.g., the coupled infrastructure of power stations and internet communication
servers \cite{IN}. In recent years, the possibility of cascades of faults through coupled interdependent networks has been pointed out as a crucial aspect with respect to the assessment and design of critical infrastructures \cite{PANEL}.}

In this paper, we consider that 
a set of identical dynamical systems $\dot{x}_i=F(x_i(t))$, $i=1,2,...,N$, are coupled through the connections of $m$ different
networks, and we refer to such a system as a \emph{hypernetwork}, see e.g., \cite{HNB2, HNB,HN}, \footnote{Another definition used in the literature to refer to such systems is that of \emph{multislice networks} \cite{multislice}.}.
We first consider the case of $m=2$ networks (a generalization to the case of $m \geq 2$ networks will be presented in Sec. IV). The systems are then coupled as follows,
\begin{equation}\label{AB}
\dot{x}_i(t)=F(x_i(t)) +  \sigma^A \sum_{j=1}^N A_{ij} [G(x_j(t-\tau_g))-G(x_i(t-\tau_g))] +  \sigma^B \sum_{j=1}^N B_{ij} [H(x_j(t-\tau_h))-H(x_i(t-\tau_h))],
\end{equation}
$i=1,2,...,N$, where $x_i(t)=[x_i^1(t),x_i^2(t),...,x_i^n(t)]^T$ is the $n$-dimensional state of node $i$, $F:R^n \rightarrow R^n$ represents the dynamics of each individual unit, $G:R^n \rightarrow R^n$ and $H:R^n \rightarrow R^n$ are different coupling functions, $\tau_g$ and $\tau_h$ are (possibly) different interaction delays, $\sigma^A$ and $\sigma^B$ are two scalar coefficients.
As can be seen from (\ref{AB}), the interactions between the individual units are those of two distinct networks, which are represented by the two distinct adjacency matrices $A=\{A_{ij}\}$ and $B=\{B_{ij}\}$. Thus Eqs. (\ref{AB}) describe a hypernetwork of coupled dynamical systems.


An equivalent way of writing Eqs. (\ref{AB}) is the following,
\begin{equation}
\dot{x}_i(t)=F(x_i(t))+\sigma^A \sum_{j=1}^N L^A_{ij} G(x_j(t-\tau_g)) +\sigma^B \sum_{j=1}^N L^B_{ij} H(x_j(t-\tau_h)), \label{LALB}
\end{equation}
$i=1,2,...,N$, where $L^A_{ij}=A_{ij}-\delta_{ij} \sum_j A_{ij}$ and  ${L}^B_{ij}= B_{ij} -\delta_{ij} \sum_j B_{ij}$ are two Laplacian matrices. 
Say $\{\lambda^A_i\}$  and $\{\lambda^B_i\}$ the set of eigenvalues associated respectively with the two matrices $L^A$ and $L^B$. By construction, both matrices $L^A$ and $L^B$ have one eigenvalue,  $\lambda^A_N=0$ and $\lambda^B_N=0$, with associated eigenvector $[1,1,...,1]$. 
{The $nN$ dimensional state space of the system in Eqs. (\ref{LALB})  contains an $n$-dimensional synchronization manifold $\mathcal{I}$},
\begin{equation}
x_1(t)=x_2(t)=...=x_N(t). \label{ss}
\end{equation}
{Note that if a solution belongs to $\mathcal{I}$ over a time interval $[t_0,t_0+\tau_{max}]$, where $\tau_{max} =\max_{\tau_g,\tau_h}$, then the solution will belong to $\mathcal{I}$, for any time $t>t_0+\tau_{max}$.  In this case, the synchronized solutions $x_1(t)=x_2(t)=...=x_N(t)=x_s(t)$ is characterized by the same dynamics as that of an uncoupled system,}
\begin{equation}
\dot{x}_s(t)=F(x_s(t)).\label{es}
\end{equation}


The main goal of this paper is to study linear stability of the synchronous solution (\ref{ss},\ref{es}) for the set of equations (\ref{LALB}).
The same problem for the case that the systems are coupled through the connections of only one network, i.e., $L^B_{ij}=0$ in Eq. (\ref{LALB}) has been intensively studied in the literature, see e.g., \cite{Pe:Ca,Wang:Chen02,Lu:Ch04,Lu:Chen:Cheng,Decoupl,Bocc1,So:di:Bo,Ki:En:Re:Zi:Ka,Pecora2009}. For this case it can be shown that linear stability of the synchronous solution can be analyzed in terms of the following low-dimensional equation
\begin{equation} \label{PC}
\delta \dot{\bar{x}}(t)= DF(x_s(t)) \delta {\bar{x}}(t) +  \sigma^A \lambda^A_k DH(x_s(t-\tau_h)) \delta {\bar{x}}(t-\tau_h),
\end{equation}
where $DF$ ($DH$) represents the Jacobian matrix of the function $F$ ($H$). In particular, the condition for stability is that the maximum Lyapunov exponents \footnote{For $\tau_h>0$, each one of the equations in (\ref{PC}) is infinite dimensional and therefore has an infinite number of Lyapunov exponents; yet there must be one among these that is the maximum.} associated with Eq. (\ref{PC}) are negative for $k=1,...,(N-1)$. Eq. (5) for $k=N$ yields
\begin{equation}
\delta \dot{\bar{x}}(t)=DF(x_s(t))  \delta {\bar{x}}(t),
\end{equation}
which corresponds to the linearized equation for the evolution in the synchronization manifold (\ref{ss}).
(\ref{PC}) is a system of  $n$ scalar differential equations as opposed to the linearized system (\ref{LALB}), which
is described by $nN$ scalar differential equations. Hence, system (\ref{PC}) is termed low-dimensional.
The nice thing about this approach is that it provides necessary and sufficient conditions for synchronization. Similar conditions have been obtained for networks of groups \cite{NSG}, for adaptive synchronization of complex networks \cite{SOTT,SAS}, for the pinning control problem applied to a complex network \cite{PC,Sorr}, and for the case that slight deviations from nominal conditions are present \cite{restr_bubbl,Su:Bo:Ni,SOPO}.  In this paper, we attempt at obtaining a condition in terms of a low-dimensional equation for the more complex case that the systems are coupled through the connections of two different networks (Eq. (\ref{LALB})).
However, as we will see, our proposed problem is not easy  to solve in general.

In what follows, we first consider the case that the two matrices $A$ and $B$ in (\ref{AB}) are arbitrary and we show that the stability problem does not admit a solution in a low-dimensional form.
Then we focus on three examples of interest for which we show that such a reduction is possible:
\begin{itemize}
  \item The two Laplacian matrices $L^A$ and $L^B$ commute.
  \item One of the networks (either $A$ or $B$) is unweighted and fully connected.
  \item One of the two networks (say e.g., $A$) is such that $A_{ij}=a_j$, $i,j=1,...,N$.
\end{itemize}

The rest of the paper is organized as follows. In Sec. II we attempt at obtaining necessary and sufficient conditions for stability of the synchronous solution for a hypernetwork (\ref{LALB}). Yet, we show that unfortunately it is not always possible to reduce the problem in a low-dimensional form. However, we analyze three cases of interest for which such a reduction is possible. Furthermore, we define a class of networks such that if one of the two coupling networks belongs to this class, the reduction can be obtained independently of the other network. Numerical simulations are shown in Sec. III. 
In Sec. IV we generalize our results to the case of hypernetworks formed of $m\geq2$ networks. 
{A more general class of hypernetworks that are not described by the set of equations (\ref{AB}) is discussed in Sec. V, where the example of a network of neurons connected by both electrical gap-junctions and chemical synapses is presented.} Finally, the conclusions are given in Sec. VI.

\section{Stability Analysis}

We consider stability of Eqs. (\ref{LALB}) about the synchronous solution (\ref{ss}). Linearization of Eqs. (\ref{LALB}) about (\ref{ss}) yields,
\begin{equation}\label{lin}\begin{split}
\delta \dot{x}_i(t) =  DF(x_s(t)) \delta {x}_i(t) & +  \sigma^A \sum_{j=1}^{N} L^A_{ij} DG(x_s(t-\tau_g)) \delta {x}_j(t-\tau_g) \\ & + \sigma^B \sum_{j=1}^{N} {L}^B_{ij}  DH(x_s(t-\tau_h)) \delta {x}_j(t-\tau_h),
\end{split}
\end{equation}
$i=1,2,...,N$. The set of equations (\ref{lin}) can be rewritten in vectorial form as follows,
\begin{equation} \label{lin2}
\begin{split}
\delta \dot{x}(t) = I_N \otimes DF(x_s(t)) \delta {x}(t)  &+   \sigma^A L^A \otimes DG(x_s(t-\tau_g)) \delta {x}(t-\tau_g) \\
& + \sigma^B {L^B} \otimes DH(x_s(t-\tau_h)) \delta {x}(t-\tau_h),
\end{split}
\end{equation}
where $\delta x(t)=[\delta x_1(t)^T, \delta x_2(t)^T,...,\delta x_N(t)^T]^T$ and the symbol $\otimes$ indicates the direct product or Kronecker product. Now we proceed under the assumption that at least one of the two Laplacian matrices, say $L^A$, is diagonalizable, i.e., $L^A=V\Lambda^A V^{-1}$, where $\Lambda^A$ is a diagonal matrix with the elements on the main diagonal being the eigenvalues $\lambda^A_1,\lambda^A_2,...,\lambda^A_N$ and $V$ is a matrix whose columns are the associated eigenvectors, 
$v_1,v_2,...,v_N$. Then, by introducing the change of variable, $\eta(t)=V^{-1} \otimes I_n \delta{x}(t)$, Eq. (\ref{lin2})   becomes,
\begin{equation}
\begin{split}
\dot{\eta}(t) = I_N \otimes DF(x_s(t)) \eta(t) & + \sigma^A  \Lambda^A \otimes DG(x_s(t-\tau_g)) \eta(t-\tau_g) \\ &+ \sigma^B \Xi \otimes DH(x_s(t-\tau_h)) \eta(t-\tau_h),   \label{H5}
\end{split}
\end{equation}
where the matrix $\Xi= V^{-1} L^B V$. It would be nice if the matrix $\Xi$ were diagonal but unfortunately there is no guarantee that this will be the case in general. Then we see from Eq. (\ref{H5}) that, different from the classical master stability function derivation \cite{Pe:Ca}, it is not possible to decouple Eq. (\ref{H5}) in $N$  blocks, each one independent of the others.

\subsection{The case that the two matrices $L^A$ and $L^B$ commute}

\begin{figure}
\centering
\includegraphics[width=3.5in]{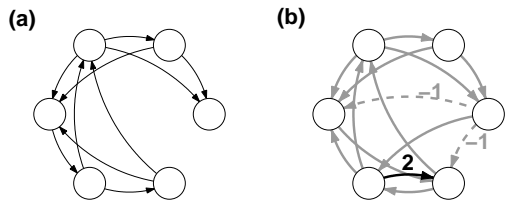}
\caption{An example of two graphs with associated commuting Laplacian matrices. (a) All the links have associated weight equal one. (b) All the links have associated weight equal one except for the link in black having associated weight 2 and the links represented as dashed arrows having associated weight -1. }
\label{Grafo}
\end{figure}

A special case is when the two matrices $L^A$ and $L^B$ commute. Two matrices that commute have the property of sharing the same set of eigenvectors, i.e., assuming that they are both independently diagonalizable, it is possible to write $L^A= V \Lambda^A V^{-1}$ and $L^B= V \Lambda^B V^{-1}$, where $\Lambda^B$ is a diagonal matrix with the elements on the main diagonal being the eigenvalues of the matrix $L^B$. Thus for this case the matrix $\Xi$ coincides with the diagonal matrix $\Lambda^B$ as $\Xi= V^{-1} V \Lambda^B V^{-1} V= \Lambda^B$. It follows that equation (\ref{H5}) can be decomposed in $N$ blocks independent of each other,
\begin{equation} \label{bocc}
 \dot{\eta}_k(t)= DF(x_s(t)) \eta_k(t)  + \sigma^A \lambda^A_k DG(x_s(t-\tau_g)) \eta_k(t-\tau_g) + \sigma^B \lambda^B_k DH(x_s(t-\tau_h)) \eta_k(t-\tau_h),
\end{equation}
$k=1,...,N$, where $\lambda^A_k$ and $\lambda^B_k$ are respectively the (complex) eigenvalues of the matrices $L^A$ and $L^B$, which are associated with the same  eigenvectors, i.e., such that $L^A v_k= \lambda^A_k v_k$ and $L^B v_k= \lambda^B_k v_k$. Recall that the eigenvalues $\lambda^A_N=\lambda^B_N=0$ and the corresponding  eigenvector is $[1,1...1]$. Then for $k=N$, Eq. (\ref{bocc}) yields,
\begin{equation} \label{boccs}
 \dot{\eta}_N(t)=DF(x_s(t)) \eta_N(t),
\end{equation}
 which corresponds to perturbations in the direction tangent to the synchronization manifold (\ref{ss}) and as such are not relevant in determining stability of the synchronous solution. Thus a necessary and sufficient condition for synchronization is that the Lyapunov exponents associated with Eq. (\ref{bocc}) are negative for $k=1,2,...,(N-1)$.

We now introduce a parametric equation
\begin{equation} \label{param1}
\dot{\eta}(t)= DF(x_s(t)) \eta(t) + y DG(x_s(t-\tau_g)) \eta(t-\tau_g) + z DH(x_s(t-\tau_h)) \eta(t-\tau_h),
\end{equation}
where $y$ and $z$ are two complex parameters.
We associate a master stability function with Eq. (\ref{param1}),
\begin{equation} \label{MSF}
\mathcal{M}(y,z),
\end{equation}
 which returns the maximum Lyapunov exponent of Eq. (\ref{param1}) as a function of the pair of complex arguments $(y,z)$.
Then given any hypernetwork (\ref{LALB}), stability of the synchronous solution can be evaluated by checking that $\mathcal{M}(y,z)<0$, for $(y,z)=(\sigma^A \lambda^A_k,\sigma^B \lambda^B_k)$, $k=1,2,...,(N-1)$. Alternatively,  a necessary and sufficient condition for stability of the synchronized evolution is that the pairs $(\sigma^A \lambda^A_k, \sigma^B \lambda^B_k)$, $k=1,2,...,(N-1)$ fall in the region of the domain of the master stability function $\mathcal{M}(y,z)$ for which $\mathcal{M}<0$.
A similar result for the case of a single network whose topology is allowed to evolve in time has been previously obtained in \cite{Boc06}.

However, we note that the case that the two matrices $L^A$ and $L^B$ commute is quite specific and not very likely to occur in practical situations. An example of two graphs with associated commuting Laplacian matrices is shown in Fig. 1.
In Sections IIB and IIC, we present two examples for which a reduction of the stability problem (\ref{lin}) in a low-dimensional form is possible, even if the two matrices $L^A$ and $L^B$ do not commute.

\subsection{The case that one of the two networks is unweighted and fully connected}

\begin{figure}
\centering
\includegraphics[width=3.5in]{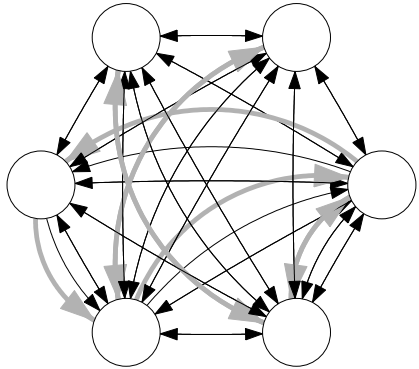}
\caption{A hypernetwork formed of a fully connected graph (thin black arrows) and a superimposed network of 9 directed links (thick gray arrows). All the links (those associated with either one of the networks) have associated weight equal one.}
\label{Grafo}
\end{figure}

We consider the case that one of the two networks is unweighted and fully connected. Without loss of generality we take this matrix to be $A$,
\begin{equation}
A_{ij}=\left\{\begin{array}{ll} 1,&  \mbox{for $i,j=1,...,N$, $j\neq i$.} \\ 0,& \mbox{for $i=j$.} \end{array} \right. \label{case2}
\end{equation}
Then $L^A_{ij}=(1-\delta_{ij}N)$, where $\delta_{ij}$ is the Kronecker delta. An example of such a hypernetwork is shown in Fig.\ 2.
We consider again stability of the synchronous solution (\ref{ss}). In what follows, we  obtain a master stability function by only diagonalizing the $(N-1)$ dimensional subspace of transverse perturbations without worrying about the fact that these may couple into the remaining direction (which is tangent to the synchronization manifold).


The matrix $L^A$ can be diagonalized as $L^A=V\Lambda^A V^{-1}$, where $\Lambda^A$ is the following diagonal matrix,
\begin{eqnarray}\Lambda^A= \{\Lambda^A_{ij} \}=
{\small\small\small{ \begin{pmatrix} - N &
0  & 0 & \cdots & 0 \cr
0 &- N & 0 & \cdots & 0   \cr
& & \ddots &  \cr
0 & 0 & \cdots & - N  & 0\cr
0 & 0 & 0 & 0 & 0
\end{pmatrix}}}. \nonumber
\end{eqnarray}

We now look at Eq. (\ref{H5}).
It can be shown that the matrix $\Xi=V^{-1} {L^B} V$, $\Xi= \{\Xi_{ij} \}$, has the form
\begin{eqnarray}\Xi= 
{\small\small\small{ \begin{pmatrix} \Xi_{11} &
\Xi_{12}  &  \cdots & \Xi_{1(N-1)} & 0 \cr
\Xi_{21} &\Xi_{22} & \cdots & \Xi_{2(N-1)} & 0   \cr
& & \vdots &  \cr
\Xi_{(N-1)1} & \Xi_{(N-1)2} & \cdots & \Xi_{(N-1)(N-1)}  & 0\cr
\Xi_{N1} & \Xi_{N2} & \cdots & \Xi_{N(N-1)} & 0
 \end{pmatrix}}}. \label{csi}
\end{eqnarray}
In fact, the matrix ${L^B} V$ has a column whose elements are all zero. This is due to the properties (i)   that the sum of the elements in each row of the matrix ${L}^B$ equals zero, and (ii) that the matrix $V$ has a column (the same column where the eigenvalue $0$ of $\Lambda^A$ is) whose elements are all the same. It immediately follows that $V^{-1} {L^B} V$ has a column whose elements are all zero.
Therefore, Eq. (\ref{H5}) can be re-expressed as,
\begin{equation}
\begin{split}
\dot{\eta'}(t) =& I_{N-1}  \otimes DF(x_s(t)) \eta'(t) \\
& - \sigma^A N   I_{N-1} \otimes DG(x_s(t-\tau_g))  \eta'(t-\tau_g)\\
& + \sigma^B \Xi' \otimes DH(x_s(t-\tau_h))  {\eta'}(t-\tau_h), \label{DEC1}
\end{split}
\end{equation}
\begin{equation}
\dot{\eta}_N(t)=DF(x_s(t))\eta_N(t)- \sigma^B DH(x_s(t)) \sum_{j=1}^{N-1} \Xi_{Nj} {\eta}_j(t-\tau_h), \label{DEC2}
\end{equation}
where the vector $\eta'=[\eta_1^T,\eta_2^T,...,\eta_{N-1}^T]^T$, and $\Xi'$ is the $(N-1)$ dimensional square matrix,
\begin{eqnarray}\Xi'= \{\Xi'_{ij} \}=
{\small\small\small{ \begin{pmatrix} \Xi_{11} &
\Xi_{12}  &  \cdots & \Xi_{1(N-1)}  \cr
\Xi_{21} &\Xi_{22} & \cdots & \Xi_{2(N-1)}    \cr
& & \vdots &  \cr
\Xi_{(N-1)1} & \Xi_{(N-1)2} & \cdots & \Xi_{(N-1)(N-1)}  \cr
\end{pmatrix}}}. \nonumber
\end{eqnarray}
We note that Eq. (\ref{DEC1}) is independent from Eq. (\ref{DEC2}). Hence, we term the first as the drive system and the second as the response system.
Note that $\eta'$ corresponds to perturbations transverse to the synchronization manifold, while $\eta_N$ corresponds to perturbations within the synchronization manifold. Thus synchronization stability is governed by Eq. (\ref{DEC1}), which does not involve $\eta_N$. We diagonalize the matrix $\Xi'$, obtaining $(N-1)$ blocks of the form,
\begin{equation} \label{H6} \begin{split}
\dot{\zeta}_k(t) =  DF(x_s(t)) \zeta_k(t) &-  \sigma^A N  DG(x_s(t-\tau_g))   \zeta_k(t-\tau_g)\\ &+ \sigma^B \nu_k DH(x_s(t-\tau_h))   {\zeta}_k(t-\tau_h),
\end{split} \end{equation}
$k=1,...,(N-1)$, where $(\nu_1,\nu_2,...,\nu_{N-1})$ are the eigenvalues of the matrix $\Xi'$. Note that the eigenvalues of the matrix $\Xi'$ are the same as those of the matrix ${L}^B$, except for the one eigenvalue $\lambda^B_N$ that is equal to $0$. 

If  the $(N-1)$ maximum Lyapunov exponents associated with the drive system  (\ref{H6}) are all negative, then for large enough $t$, $\zeta_k(t) \rightarrow 0$, $k=1,...,(N-1)$. If this happens, then  Eq. (\ref{DEC2}) yields for large enough $t$,
\begin{equation}
\dot{\eta}_N(t)=DF(x_s(t))\eta_N(t),
\end{equation}
which corresponds to the linearized equation in the direction tangent to the synchronization manifold. 

Thus we can introduce the parametric equation (\ref{param1}) in the pair ($y$, $z$), with the parameter $z$ being possibly complex
and a master stability function (\ref{MSF}) which returns the maximum Lyapunov exponent
of Eq. (\ref{param1}) as a function of the parameters $y$ and $z$. For a given hypernetwork (\ref{LALB},\ref{case2}), a necessary and sufficient condition for stability of the synchronous solution (\ref{ss}), is that $y=-\sigma^A N$ and $z=\sigma^B \nu_k$, $k=1,2,...,(N-1)$ belong to the region of the domain of the master stability function (\ref{MSF}), for which $\mathcal{M}(y,z)<0$.

This formulation allows to decouple the effects of the dynamical function $F$ and the coupling functions $G$ and $H$,
from those of the  network matrices $L^A$ and $L^B$. In particular, for any given triplet of functions $F,G$, and $H$, the matrix $B$ determines the parameters $\nu_1,\nu_2,...,\nu_{N-1}$, and if the master stability function $\mathcal{M}(y,z)$ is negative for $y=-\sigma^A N$ and $z=\sigma^B \nu_1, \sigma^B \nu_2,..., \sigma^B \nu_{(N-1)}$, then the synchronization manifold is stable.
 An interesting thing about our derivation (\ref{H6}) is that we have been able to obtain a reduction of the stability problem (\ref{lin}) in a low-dimensional form though the two matrices $L^A$ and $L^B$ do not necessarily commute.

\subsection{The case that $A_{ij}=a_j$}

\begin{figure}
\centering
\includegraphics[width=5in]{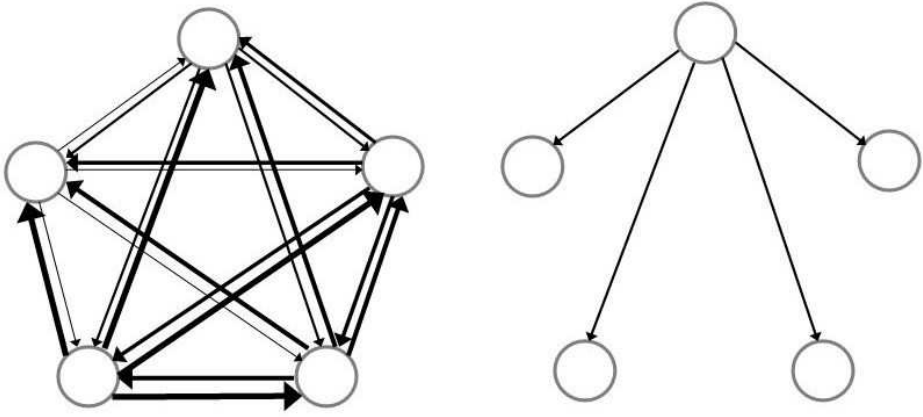}
\caption{On the left-hand-side, a $N=5$-node network belonging to class $\mathcal{C}$, i.e., such that the entries of the associated adjacency matrix $A=\{A_{ij}\}$ satisfy $A_{ij}=a_j$ (the condition discussed in Sec. IIC). The width of each link $j \rightarrow i$ represents the strength of the associated coupling $A_{ij}$. The network on the right-hand-side is an \emph{outward star graph}, corresponding to satisfying (\ref{aij}) with $a_j=0$, $j=1,...,(N-1)$.}
\label{Pent}
\end{figure}

Here we consider the case that the coupling strength from node $j$ to node $i$ is only a function of the \emph{source node} $j$ and not of the \emph{destination node} $i$, that is
\begin{equation} \label{aij}
A_{ij}=a_j, \quad i,j=1,...,N.
\end{equation}
{An example of such a network is shown on the left-hand-side of Fig.\ \ref{Pent}, where the width of each link $j \rightarrow i$ represents the strength of the associated coupling $A_{ij}$. The network on the right-hand-side of Fig.\ \ref{Pent} is an \emph{outward star graph}, corresponding to setting $a_j=0$ for $j=1,...,(N-1)$ and $a_N\neq0$ in Eq.\ (\ref{aij}).}
Under assumption (\ref{aij}), Eqs. (\ref{AB}) become,
\begin{align}\label{AB2}
\dot{x}_i(t)=F(x_i(t)) + & \sigma^A \sum_{j=1}^N a_j [G(x_j(t-\tau_g))-G(x_i(t-\tau_g))] \\ + & \sigma^B \sum_{j=1}^N B_{ij} [H(x_j(t-\tau_h))-H(x_i(t-\tau_h))], \nonumber
\end{align}
$i=1,2,..,N$, which can be recast in the form of Eq. (\ref{LALB}), with the matrix $L^A= \{L^A_{ij} \}$ having the form,
\begin{eqnarray}L^A=
{\small\small\small{ \begin{pmatrix} a_1-\bar{a} &
a_2  & \cdots & a_{(N-1)} & a_N \cr
a_1 & a_2-\bar{a} & \cdots & a_{(N-1)} & a_N   \cr
& & \ddots &  \cr
a_1 & a_2 & \cdots & a_{(N-1)}-\bar{a}  & a_N\cr
a_1 & a_2 & \cdots & a_{(N-1)} & a_N-\bar{a}
\end{pmatrix}}}, \label{LA}
\end{eqnarray}
where $\bar{a}=\sum_{j=1}^N a_j$. The matrix $L^A$ in (\ref{LA}) has the property that it has one eigenvalue $\lambda^A_N=0$ with associated eigenvector $[1,1,...,1]$ and the remaining $(N-1)$ eigenvalues are  $\lambda^A_1=\lambda^A_2=...=\lambda^A_{(N-1)}=-\bar{a}$. Moreover $L^A$ can be diagonalized as $L^A=V \Lambda^A V^{-1}$ with $\Lambda^A$ equal,
\begin{eqnarray}\Lambda^A= \{\Lambda^A_{ij} \}=
{\small\small\small{ \begin{pmatrix} -\bar{a} &
0  & 0 & \cdots & 0 \cr
0 &-\bar{a} & 0 & \cdots & 0   \cr
& & \ddots &  \cr
0 & 0 & \cdots & -\bar{a}  & 0\cr
0 & 0 & 0 & 0 & 0
\end{pmatrix}}}. \nonumber
\end{eqnarray}
It is easy to see that the matrix $\Xi$ is in the form (\ref{csi}), with the entries in the $N$-column being all equal to zero. This allows  to decouple the  set of linearized equations into a drive subsystem and a response subsystem, with the response subsystem corresponding to perturbations tangent to the synchronization manifold (\ref{ss}) and the drive subsystem corresponding to perturbations transverse to the synchronization manifold.

Then following the analysis in Sec. IIB, it can be shown that a necessary and sufficient condition for stability of the synchronous solution for the hypernetwork (\ref{AB2}) is that the maximum Lyapunov exponent of the low-dimensional equation,
\begin{equation} \label{LD3}
\dot{\theta}_k(t) =  DF(x_s(t)) {\theta}_k(t) - \sigma^A \bar{a}  DG(x_s(t-\tau_g)) {\theta}_k(t-\tau_g) +\sigma^B \lambda^B_k DH(x_s(t-\tau_h))  {\theta}_k(t-\tau_h),
\end{equation}
is negative for $k=1,...,(N-1)$, where $\lambda^B_1,\lambda^B_2,...,\lambda^B_{(N-1)}$ are the eigenvalues of the matrix $L^B$, excluding the one eigenvalue $\lambda^B_N=0$. It is then possible to associate Eq. (\ref{LD3}) with the parametric equation (\ref{param1})
and the master stability function (\ref{MSF}), which returns the maximum Lyapunov exponent of Eq. (\ref{param1}) as a function of the pair of parameters $(y,z)$, with the parameter $y=- \sigma^A \bar{a}$ and the (possibly complex) parameter $z=\sigma^B \lambda^B_1, \sigma^B \lambda^B_2,..., \sigma^B \lambda^B_{(N-1)}$. Again we note that we have been able to obtain a reduction of the stability problem (\ref{lin}) in a master stability function form though the two matrices $L^A$ and $L^B$ do not necessarily commute.

We wish to emphasize that the case in Sec. IIB (fully connected network) can be seen as a subcase of that in Sec. IIC ($A_{ij}=a_j$). In fact, if we assume $a_j=a$, $j=1,...,N$ in (\ref{aij}), then the Laplacian matrix $L^A= \{L^A_{ij} \}$ in (\ref{LA}) is such that $L^A_{ij}=a(1-\delta_{ij}N)$, i.e., it coincides with the matrix $L^A$ considered in Sec. IIB up to a multiplicative factor $a$.

\subsection{Necessary conditions on the matrix $A$}

We observe here that there is a substantial difference between the conditions on the adjacency matrices $A$ and $B$ (the Laplacian matrices $L^A$ and $L^B$) discussed in Sec. IIA and those discussed in Secs. IIB and IIC. First consider  the case presented in Sec. IIA, that the two Laplacian matrices $L^A$ and $L^B$ commute; then, if one of the two matrices changes, there is no guarantee that the condition would still hold. On the other hand, the conditions discussed in Sec. IIB and IIC refer essentially to one of the two matrices, allowing the other one to be freely chosen. 

In sections IIB and IIC, we have found sufficient conditions on one of the two adjacency matrices, say $A$, that if satisfied, allow a reduction of the stability problem in a low dimensional form, irrespective of the other adjacency matrix, say $B$. In this section, we are interested in finding necessary condition for this to happen. 
We consider the set of Eqs. (\ref{AB}) and we define  the class $\mathcal{C}$ of all the networks $A$ that satisfy the property of allowing a reduction of the stability problem in a low dimensional form, irrespective of the other network $B$.
In what follows, we show that a network in $\mathcal{C}$ is such that the entries of the associated adjacency matrix $A=\{A_{ij}\}$ satisfy $A_{ij}=a_j$, i.e., the same condition discussed in Sec. IIC.


Hereafter, we seek to find the conditions for an adjacency matrix $A$ (a Laplacian matrix $L^A$) to be in $\mathcal{C}$.
Based on our previous discussion in Secs. IIB and IIC, we see that the properties that the matrix $L^A$ has to satisfy are the following:

(A)   $L^A$ is diagonalizable.

(B) The sums of the elements in the rows of the matrix $L^A$ are equal zero. This also implies that the matrix $L^A$ has one eigenvalue equal zero, with associated eigenvector $[1,1,...,1]$.

(C) The remaining $(N-1)$ eigenvalues are all the same.

If the three properties above are satisfied, the matrix $L^A$ can  always be written as follows,
\begin{equation}\label{WPW}
L^A=W P W^{-1},
\end{equation}
where the matrix $P$ is a diagonal matrix with all the entries on the main diagonal being equal to the same value, say $p$, except one entry (which, without loss of generality, we assume to be the one in the rightmost column) that is equal to zero. The matrix $W$ is any invertible matrix with the rightmost column being equal to the vector $[1,1,...,1]$. We note that the matrix $P$ can be rewritten as $P=p (I_N-I_N^*)$, where $I_N$ is the identity matrix and $I_N^*$ is a diagonal matrix with all the entries on the main diagonal being equal to zero except the one in the rightmost column being equal to one. It follows that
\begin{equation}
L^A=p (I- W I_N^* W^{-1}).
\end{equation}
It is easy to see that the matrix $W I_N^* W^{-1}$ is by construction such that the entries in each one of its columns are the same. Hence, the corresponding adjacency matrices $A$ have to be in the form $A_{ij}=a_j$, discussed in Sec. IIC.

We conclude that if we are given a specific adjacency matrix $B$ (a specific Laplacian matrix $L^B$), there are two possible choices of the adjacency matrix $A$ (the Laplacian matrix $L^A$) for which the stability problem can be reduced in a low-dimensional form: (i) $L^A$ commutes with $L^B$, and (ii) $A$ belongs to $\mathcal{C}$, i.e., its entries are such that  $A_{ij}=a_j$. Note that condition (ii) is independent of the choice of the matrix $L^B$. 

\section{Examples}

\subsection{Example 1: Coordinated motion of swarms of particles.}

 Swarms of birds, hordes of insects, shoals of fishes, and colonies of ants have been modeled as systems of interacting self-propelled particles \cite{OSFLOCK,Cu:Sm,Ab:Po}. Here we consider a simple model of $N$ particles moving along a fixed direction, say $y$, through a resistent fluid.  The position (velocity) of particle $i$ along the $y$ direction is labeled as $y_i(t)$ ($v_i(t)$), $i=1,...,N$. We consider the following equations of motion,
\begin{subequations}\label{sw}
\begin{align}
\dot{y}_i(t)= & v^r_i(t), \\
\dot{v}_i(t)= & (\alpha-\beta v^r_i(t)^2) v^r_i(t) +\sum_j m_j (y_j(t)-y_i(t)) +\sum_j m_j c_{ij}(t) (v_j(t)-v_i(t)),
\end{align}
\end{subequations}
$i=1,...,N$. The first term on the right hand-side of Eq. (\ref{sw}b) represents propulsion/friction of particle $i$, $v^r_i(t)$ is the relative velocity along $y$  with respect to the resistent fluid of particle $i$, $v^r_i(t)=(v_i(t)-v_f(t))$ and $v_f(t)$ is the velocity of the resistent fluid, which we model as an external input and we assume to be uniform in space. The second term on the right hand-side of Eq. (\ref{sw}b) represents attraction from particle $j$ on particle $i$. The third term on the right hand-side of Eq. (\ref{sw}b) models a relative velocity adjustment between  particles.  $m_j> 0$ is the \emph{mass} of particle $j=1,...,N$, $\alpha,\beta \geq 0$, $c_{ij}(t)$ measures the strength of the interaction from particle $j$ on particle $i$, which we set to be a function of the physical distance between particles $i$ and $j$,
\begin{equation}
c_{ij}(t)=[d_{ij}^2+(y_j(t)-y_i(t))^2]^e,
\end{equation}
where $d_{ij}$ is the distance between particles $i$ and $j$ in the plane orthogonal to the $y$ direction and the exponent $e$ determines the strength of the interaction as a function of the distance. An analogous model for particles that are allowed to move in the three-dimensional space has been considered in \cite{Ca:Ca:Ro}. 

We note that the system of equations (\ref{sw}) admits a synchronous solution $y_1(t)=y_2(t)=...=y_N(t)=y_s(t)$, $v_1(t)=v_2(t)=...=v_N(t)=v_s(t)$, obeying
\begin{subequations}\label{sws}
\begin{align}
\dot{y}_s(t)= & v_s(t), \\
\dot{v}_s(t)= & [\alpha-\beta (v_s(t)-v^f(t))^2] (v_s(t)-v^f(t)),
\end{align}
\end{subequations}
where again $v^f(t)$ is an external input.
The synchronous solution corresponds to a configuration in which all the positions and the velocities of the particles along the $y$-direction are the same. 
We are interested in studying stability of this solution. In order to do that, we linearize Eq. (\ref{sw}) about (\ref{sws}),
\begin{subequations}\label{lsw}
\begin{align}
\delta \dot{y}_i(t)= & \delta v_i(t), \\
\delta \dot{v}_i(t)= & [\alpha-3 \beta (v_s(t)-v^f(t))^2] \delta v_i(t) +\sum_j m_j (\delta y_j(t)- \delta y_i(t)) +\sum_j m_j (d_{ij})^{2e} (\delta v_j(t)- \delta v_i(t)),
\end{align}
\end{subequations}
$i=1,...,N$.
Equations (\ref{lsw}) can be rewritten in matrix form,
\begin{equation}
\begin{split}
\delta \dot{x}_i(t) = \left(
  \begin{array}{cc}
    0 & 1 \\
    0  & [\alpha-3 \beta (v_s(t)-v^f(t))^2] \\
  \end{array}
\right) \delta {x}_i(t)  +   \left(
  \begin{array}{c}
    1 \\
    0 \\
  \end{array}
\right) \sum_j A_{ij} [\delta {x}_j(t)-\delta x_i(t)] +   \left(
  \begin{array}{c}
    0 \\
    1 \\
  \end{array}
\right) \sum_j B_{ij} [\delta {x}_j(t)-\delta x_i(t)],
\end{split}
\end{equation}
where
\begin{equation}
\delta {x}_i(t)=\left(
  \begin{array}{c}
    \delta {y}_i(t) \\
    \delta {z}_i(t)) \\
  \end{array}
\right),
\end{equation}
and
$A_{ij}=m_j$, $B_{ij}=m_j (d_{ij})^{2e}$, $i,j=1,...,N$. It is easy to see that the matrix $A=\{A_{ij}\}$ belongs to class $\mathcal{C}$. Hence, following Sec. IIC, the stability problem can be reduced in a low-dimensional form analogous to Eq. (\ref{LD3}),
\begin{equation}
\dot{\theta}_k(t) = \left(
  \begin{array}{cc}
    - \bar{a} & 1 \\
    0  & [\alpha-3 \beta (v_s(t)-v^f(t))^2] +\lambda^B_k \\
  \end{array}
\right)     {\theta}_k(t),
\end{equation}
$k=1,...,(N-1)$. Note that $\bar{a}=\sum_{j=1}^N m_j>0$.
Thus a necessary and sufficient condition for the synchronous solution to be stable is that $<(v_s(t)-v^f(t))^2>_t >(\alpha+\lambda^B_k)/(3 \beta)$, $k=1,...,(N-1)$, where again we have used the symbol $<...>_t$ to indicate the time-average.

\subsection{Example 2: Synchronized Chaotic Motion.}

In what follows, we consider a hypernetwork that allows a chaotic synchronous evolution (\ref{es}).
 We choose $n=3$,
\begin{equation}
F({{x}})=\left[\begin{array}{c}
    -x_{2}-x_{3}  \\
    x_{1} + 0.2 x_{2} \\
    0.2+ (x_{1}-7 )  x_{3}
  \end{array} \right], \label{RC}
\end{equation}
is the equation of the chaotic R\"ossler oscillator, $G(x(t))=[x_1(t),0,0]^T$, and $H(x(t))=[0,x_2(t),0]^T$, $\tau_g=\tau_h=0$. 
{Stability of the synchronous solution for networks of coupled R\"ossler oscillators coupled via different coupling functions has been widely investigated in the literature, see e.g., \cite{Pecora2009}. While it is known that this problem allows a low-dimensional reduction, the case of dynamical hypernetworks has not been considered. In what follows, we show that under specific conditions, a low-dimensional analysis can still be applied and based on this approach, we derive new conditions for the  stability of the synchronous solution. We numerically compute the master stability function $\mathcal{M}(y,z)$ associated with Eq. (\ref{param1}) for the case that $y$ and $z$ are real numbers.}

Fig.\ 4 shows the results of our computations with the gray (white) area indicating a negative (positive) master stability function. We wish to emphasize that once the master stability function has been computed for a given triplet $F$, $G$, and $H$ (as shown in Fig.\ 4), we are able to predict stability of the synchronous solution for any hypernetwork (\ref{LALB}), corresponding to either one of the three cases presented in Sections IIA, IIB, and IIC.

\begin{figure}
\centering
\includegraphics[width=3in]{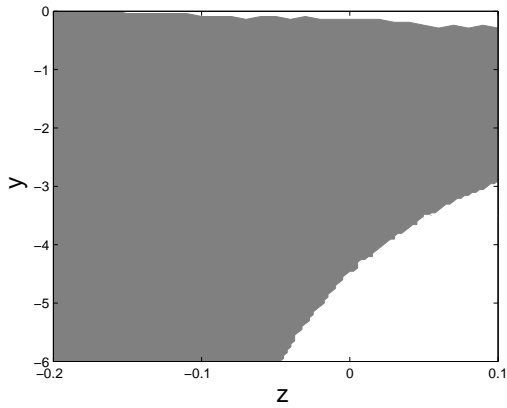}
\caption{Sign of the master stability function $\mathcal{M}(y,z)$ for a network of R\"ossler systems (\ref{RC}), $G(x(t))=[x_1(t),0,0]^T$ and $H(x(t))=[0,x_2(t),0]^T$. The gray (white) area indicates a negative (positive) maximum Lyapunov exponent. }
\label{fig_sim}
\end{figure}

\begin{figure}
\centering
\includegraphics[width=3.1in]{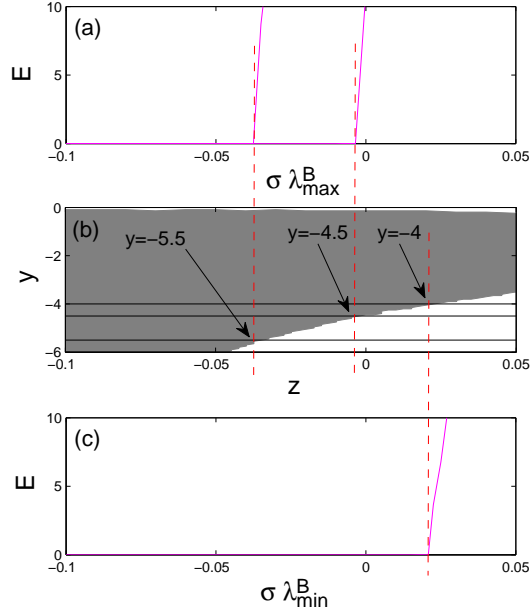}
\caption{(b) shows the intersections between  the right profile of the stability area of Fig.\ 3 and the lines $y=-4$, $y=-4.5$, and $y=-5.5$. (a) and (c) show the results of numerical simulations for which we have have integrated Eqs. (\ref{LALB}) and (\ref{RC}) with $G(x(t))=[x_1(t),0,0]^T$ and $H(x(t))=[0,x_2(t),0]^T$ for a long time and recorded the average synchronization error $E$. }
\label{fig_sim}
\end{figure}
We define the average synchronization error $E$,
\begin{equation}
{E}=({Nn})^{-1} { \sum_{i=1}^N \sum_{\ell=1}^n { \rho_\ell^{-1}<|x_{i\ell}(t)-\bar{x}_{i\ell}(t)|>_t}},\label{error}
\end{equation}
where $\bar x_{i\ell}(t)=N^{-1} \sum_{i=1}^{N} x_{i\ell}(t)$, $\rho_\ell=$ $<(x_{s\ell} - <x_{s\ell}>)^2>^{1/2}$,  $<...>_t$ indicates a time average and $x_s=(x_{1s},x_{2s},x_{3s})^T$ denotes the dynamics of an uncoupled system (i.e., using dynamics from Eq. (\ref{es})).

We consider the hypernetwork shown in Fig.\ 2. We assume that the matrix $A$ is associated with the fully connected network (whose connections are represented as thin black arrows in the figure) and that the matrix $B$ is associated with the superimposed graph (in gray in the figure), i.e., the entries of the matrix $B$ are $B_{ij}=1$ if there is a direct arrow from node $j$ to node $i$ in the figure, $B_{ij}=0$ otherwise. Then we have that the  eigenvalues of the matrix $L^B$ are  $\{-3,  -2.618,  -2,   -1,   -0.382,  0\}$; note that they are all real and less or equal zero. In general, in order to verify stability, it is necessary to check that all the pairs  $(y,z)=(\sigma^A \lambda^A_k,\sigma^B \lambda^B_k)$, $k=1,2,...,(N-1)$  follow into the domain of the master stability function for which ${\mathcal{M}}(y,z)<0$. This can be done for example by superimposing the  $(N-1)$ points corresponding to all the pairs $(\sigma^A \lambda^A_k,\sigma^B \lambda^B_k)$ to Fig.\ 4; if all the points fall into the grey area, this ensures stability (sufficient condition for synchronization) and if only one of the points fall into the white area, this corresponds to instability (necessary condition for synchronization).
However, for the network of Fig.\ 2 and the master stability function of Fig.\ 4,  we note that for any fixed value of $y=-\sigma^A N$, the condition for stability is that
\begin{equation} \label{conditio}
\sigma^B \lambda^B_i< \kappa, \qquad i=1,...,(N-1),
\end{equation}
where the parameter $\kappa$ is the abscissa of the intersection of the $y=-\sigma^A N$ line with the right profile of the gray area shown in Fig.\ 4. Note that $\kappa$ can be either positive or negative. We define $\lambda^B_{max}=\max {(\lambda^B_1,\lambda^B_2,...,\lambda^B_{(N-1)})}$ and $\lambda^B_{min}=\min{(\lambda^B_1,\lambda^B_2,...,\lambda^B_{(N-1)})}$.
Then for this case, stability of the synchronous solution can be assessed by testing the following simple condition,
\begin{subequations} \label{conditiones}
\begin{align}
\begin{array}{ll} \sigma^B \lambda^B_{max}<\kappa,&  \mbox{if $\kappa<0$,} \\
\sigma^B \lambda^B_{min}<\kappa,& \mbox{if $\kappa>0$.} \end{array} \nonumber
\end{align}
\end{subequations}


In Fig.\ 5 we consider the following three cases: $\sigma^A=4.5/6$, $\sigma^A=5.5/6$, and $\sigma^A=2/3$ (corresponding respectively to, $y=-4.5$, $y=-5.5$, and $y=-4$). As can be seen from Fig.\ 5(b), for the first two cases $\kappa<0$, while for the latter $\kappa>0$. 
We integrate Eqs. (\ref{LALB}) and (\ref{RC}) with $G(x(t))=[x_1(t),0,0]^T$ and $H(x(t))=[0,x_2(t),0]^T$ for a long time and record the average synchronization error $E$. As can be seen from Fig.\ 5(a)  (5(c)), $E$ approaches zero iff $\sigma^B \lambda^B_{max}<\kappa$ when $\sigma^A=4.5/6$ and $\sigma^A=5.5/6$ ($\sigma^B \lambda^B_{min}<\kappa$ when $\sigma^A=2/3$), thus confirming the master stability functions predictions.

\section{Generalization to $m$ networks}

In this section we consider synchronization of a hypernetwork formed of $m\geq 2$ distinct networks. For this case, we rewrite  Eq. (\ref{LALB}) as follows,
\begin{equation}
\dot{x}_i(t)=F(x_i(t))+\sum_{k=1}^m \sigma^k \sum_{j=1}^N L^k_{ij} G^k(x_j(t-\tau^k)), \label{LK}
\end{equation}
$i=1,2,...,N$, where  $G^k:R^n \rightarrow R^n$ is the coupling function associated with the connections of network $k$, $L^k=\{L^k_{ij}\}$ is the Laplacian matrix associated with network $k$, $\sigma^k$ is a scalar measuring the overall coupling strength for network $k$,
$k=1,...,m$. In what follows, we will generalize the main results of Sec. II  to this more general case (Eq. (\ref{LK})). The delays $\tau^k$ may be possibly different, i.e., $\tau_i \neq \tau_j$, $i,j=1,...,m$, $i \neq j$. {The $nN$ dimensional state space of the system described by Eqs. (\ref{LK})  contains the $n$-dimensional synchronization manifold $\mathcal{I}$, defined by Eq. (\ref{ss})}.
{Note that if a solution belongs to $\mathcal{I}$ over a time interval $[t_0,t_0+\tau_{max}]$, where $\tau_{max} =\max_i \tau^i$, then the solution will belong to $\mathcal{I}$, for any time $t>t_0+\tau_{max}$.  In this case, the synchronized solutions $x_1(t)=x_2(t)=...=x_N(t)=x_s(t)$ is characterized by the same dynamics as that of an uncoupled system (\ref{es}). In what follows, we are interested in evaluating stability of the synchronization manifold $\mathcal{I}$.

As a first case, we consider that the matrices $\{L^k\}$, $k=1,...,m$ all commute with each other, i.e., they all share the same set of linearly independent eigenvectors. Then, similar to Sec. IIA, 
it can be shown that stability of the synchronous solution can be reduced in the following low-dimensional form,
\begin{equation} \label{boccm}
\dot{\eta}_l(t)= DF(x_s(t)) \eta_l(t) + \sum_{k=1}^m \sigma^k \lambda^k_l DG^k(x_s(t-\tau^k)) \eta_l(t-\tau^k),
\end{equation}
$l=1,...,N$, where $\{\lambda^k_l\}$ is the set of (complex) eigenvalues of the matrices $\{L^k\}$, which are associated with the same eigenvectors, i.e., such that $L^k v_l= \lambda^k_l v_l$, $k=1,...,m$ and $l=1,...,N$. Recall that for any $k=1,...,m$, the eigenvalue $\lambda^k_N=0$, and the corresponding eigenvector is $[1,1...1]$. Hence, for $k=N$, Eq. (\ref{boccm}) yields Eq. (\ref{boccs}) which corresponds to perturbations in the direction tangent to the synchronization manifold (\ref{ss}) and as such are not relevant in determining stability of the synchronous solution. Thus a necessary and sufficient condition for synchronization is that the Lyapunov exponents associated with Eq. (\ref{boccm}) are negative for $k=1,2,...,(N-1)$. It is then possible to associate the following master stability function with Eq. (\ref{boccm})
\begin{equation}\label{mm}
\mathcal{M}(y^1,y^2,...,y^m),
\end{equation}
which returns the maximum Lyapunov exponent of the system (\ref{boccm}) for $y^k=\sigma^k \lambda^k_l$. A necessary and sufficient condition for stability is that $\mathcal{M}(y^1,y^2,...,y^m)<0$ for $l=1,...,(N-1)$.

We now 
attempt to generalize the result of Sec. IIC to a hypernetwork formed of $m$ networks. We assume that the first $(m-1)$ networks, $k=1,...,(m-1)$, belong to $\mathcal{C}$, while the remaining network, $k=m$, is arbitrary. Under these assumptions the first $(m-1)$ Laplacian networks are in the following form:
\begin{eqnarray}L^k=
{\small\small\small{ \begin{pmatrix} a_1^k-\bar{a}^k &
a_2^k  & \cdots & a_{(N-1)}^k & a_N^k \cr
a_1^k & a_2^k-\bar{a}^k & \cdots & a_{(N-1)}^k & a_N^k   \cr
& & \ddots &  \cr
a_1^k & a_2^k & \cdots & a_{(N-1)}^k-\bar{a}^k  & a_N^k \cr
a_1^k & a_2^k & \cdots & a_{(N-1)}^k & a_N^k-\bar{a}^k
\end{pmatrix}}}, \label{LAK}
\end{eqnarray}
where $\bar{a}^k=\sum_{j=1}^N a_j^k$, $k=1,...,(m-1)$. Note that two matrices in $\mathcal{C}$, i.e., having the form (\ref{LAK}), do not necessarily commute. Each matrix $L^k$ in (\ref{LAK}) has the property that it has one eigenvalue $\lambda^k_N=0$ with associated eigenvector $[1,1,...,1]$ and the remaining $(N-1)$ eigenvalues are  $\lambda^k_1=\lambda^k_2=...=\lambda^k_{(N-1)}=-\bar{a}^k$, $k=1,...,(m-1)$.

The eigenvectors of any of these matrices 
can be used as a new basis, say we choose $k=1$, $L^1=V \Lambda^1 V^{-1}$. Then it is easy to see that all the matrices $V^{-1} \Lambda^k V$, for $k=2,...,m$, are in the form
(\ref{csi}). It follows (similarly to Sec. IIC) that we can decouple the  set of linearized equations in a drive subsystem and a response subsystem, with the response subsystem corresponding to perturbations tangent to the synchronization manifold (\ref{ss}) and the drive subsystem corresponding to perturbations transverse to the synchronization manifold.
 Moreover, it can be shown that a necessary and sufficient condition for stability of the synchronous solution for the hypernetwork (\ref{AB2}) is that the maximum Lyapunov exponent of the low-dimensional equation,
\begin{equation}
\dot{\eta}_l(t) =  DF(x_s(t)) {\eta}_l(t) - \sum_{k=1}^{m-1} \sigma^k \bar{a}^k  DG^k(x_s(t-\tau^k)) {\eta}_l(t-\tau^k) +\sigma^m \lambda^m_l DG^m(x_s(t-\tau^m)) {\eta}_l(t-\tau^m) ,
\end{equation}
is negative for $l=1,...,(N-1)$, where $\lambda^m_1,\lambda^m_2,...,\lambda^m_{(N-1)}$ are the eigenvalues of the matrix $L^m$, excluding the one eigenvalue $\lambda^m_N=0$. A necessary and sufficient condition for stability is  that the master stability function (\ref{mm}) is negative  for $l=1,...,(N-1)$, where $y^k=- \sigma^k \bar{a}^k$, $k=1,...,(m-1)$ and $y^m=\sigma^m \lambda^m_l$, $l=1,...,(N-1)$.

Finally, we consider the more general case that $m'<m$ networks of the hypernetwork (\ref{LK}) belong to $\mathcal{C}$ and the remaining $(m-m')$ Laplacian matrices commute with each other.  Without loss of generality, we assume that the first $m'$ networks in (\ref{LK}) are in $\mathcal{C}$, $k=1,...,m'$, and that the remaining $(m-m')$ Laplacian matrices $L^k$  commute with each other, $k=(m'+1),...,m$. We observe that a reduction of the synchronization stability problem in a low dimensional form is possible,
\begin{equation}
\dot{\eta}_l(t) =  DF(x_s(t)) {\eta}_l(t) - \sum_{k=1}^{m'} \sigma^k \bar{a}^k  DG^k(x_s(t-\tau^k)) {\eta}_l(t-\tau^k)   +\sum_{k=(m'+1)}^{m} \sigma^k \lambda^k_l DG^k(x_s(t-\tau^k))   {\eta}_l(t-\tau^k),
\end{equation}
$l=1,...,N$, where $\lambda^k_1,\lambda^k_2,...,\lambda^k_{(N-1)}$ are the eigenvalues of the matrix $L^k$, $k=(m'+1),...,m$, which are associated with the same eigenvectors, i.e., such that $L^k v_l= \lambda^k_l v_l$, $k=(m'+1),...,m$ and stability of the low-dimensional equation can be associated with the master stability function (\ref{mm}), where
\begin{align}
y^k= \left\{\begin{array}{ll} - \sigma^k \bar{a}^k ,&  \mbox{$k=1,...,m'$,} \\ \sigma^k \lambda^k_l,& \mbox{$k=(m'+1),...,m$,} \end{array} \right.
\end{align}
$l=1,...,(N-1)$. The eigenvalue $\lambda^{m'+1}_N=...=\lambda^{m}_N=0$, with associated eigenvector $[1,...,1]^T$, represents perturbations tangent to the synchronization manifold and as such is not relevant in determining stability of the synchronous solution.

%
%
%
%
%

\section{Stability analysis for a more general class of hypernetworks} 

In this section, we consider hypernetworks of coupled systems, which cannot be cast into the specific form of Eqs. (\ref{AB}).
We will show that under appropriate conditions, the master stability reduction studied in Sec. II can be extended to study synchronization for this more general class of hypernetworks.  In particular, we focus on synchronization of neuronal networks. Global synchronization of large areas of the brain is usually associated with the onset of a pathological condition, such as Parkinson's disease or epilepsy \cite{EpilepsyBOOK}.

We study a hypernetwork of neurons coupled through both chemical synapses and electrical gap junctions.  Such neuronal
networks of different types connecting the same set of neurons have
recently
been explicitly discussed in the context of the C. Elegans nervous system,
which has both a gap junctional network and a chemical synaptic network \cite{SPCENN,MORCGCENS}. Following \cite{GAO,IZH,Ad:Pr:Dh}, a neuronal hypernetwork with these characteristics can be described by the following system of differential equations,
\begin{equation}\label{IAP} \begin{split}
\dot{x}_i(t)= & F(x_i(t)) \\ + &  \frac{\sigma^A}{k^A_i} (E_j - \varsigma^T x_i(t))  \sum_{j=1}^N A_{ij} s_{ij}(t) \varsigma \\  + & \frac{\sigma^B}{k^B_i} \sum_{j=1}^N B_{ij} \Gamma [x_j(t)-x_i(t)],
\end{split} \end{equation}
where the $n$-dimensional vector $x_i(t)=[x_i^1(t),x_i^2(t),...,x_i^n(t)]$ is the state of neuron $i$, with the first variable $x_i^1(t)$ representing its membrane potential, $F:R^n \rightarrow R^n$ defines the dynamics of an uncoupled neuron, the coupling matrix  $A=\{A_{ij}\}$ specifies the connection topology of the network of chemical synapses $j \rightarrow i$, while the coupling matrix $B=\{B_{ij}\}$ specifies the connection topology of the network formed of electrical gap junctions $j \leftrightarrow i$, ${k^A_i}=\sum_j A_{ij}$, ${k^B_i}=\sum_j B_{ij}$, $\sigma^A$ and $\sigma^B$ are two scalar coefficients, $E_j$ is the synaptic reverse potential of neuron $j$. Note that the matrix $A$ ($B$) is assumed to be asymmetrical (symmetrical).
The $n$-matrix
\begin{eqnarray}\Gamma=
{\small\small\small{ \begin{pmatrix} 1 &
0  & \cdots & 0 \cr
0 & 0 & \cdots & 0   \cr
& &  \ddots \cr
0 & 0 & 0 & 0
\end{pmatrix}}}. \nonumber
\end{eqnarray}
 specifies the form of the coupling, indicating that neurons are coupled through their membrane potentials, the n-vector $\varsigma=[1,0,...,0]^T$ has a similar function, that is, selecting the first state variable $x_i^1$ of the state-vector $x_i$; $\Gamma\equiv\varsigma \varsigma^T$.
The dynamical variables $s_{ij}(t)$ represent how strongly  cell $j$ is connected to cell $i$ and obey the following differential equation \cite{GAO},
\begin{equation}\label{sij}
\dot{s}_{ij}(t)=-c_1 s_{ij}(t)+c_2 (1-s_{ij}(t))S(\varsigma^T x_j(t-\tau)),
\end{equation}
$i,j=1,...,N$, where
$\tau$ is the interaction delay associated with synaptic coupling (due to axonal conduction and synaptic processes), $c_1,c_2>0$ are two scalar coefficients, $S:R \rightarrow R$ is a sigmoidal function, which we set,
\begin{equation}
 S(\varsigma^T x_j(t-\tau))=1+\tanh((\varsigma^T x_j(t-\tau)-v_{th})/v_{sl}),
 \end{equation}
 where $v_{sl}^{-1}$ represents the slope of the function $S$ when its argument is small and $v_{th}$ is the firing threshold.
As can be seen from (\ref{IAP}), the  individual neurons may simultaneously interact through two distinct networks, i.e., the network $A$ formed of chemical synapses and the network $B$ formed of electrical gap junctions.

The condition for the set of equations (\ref{IAP}) to admit a synchronous solution
\begin{subequations}\label{xxx}
\begin{align}
x_1(t)= & x_2(t)=...=x_N(t)=x_s(t), \\
s_{11}(t)= & s_{12}(t)=...=s_{NN}(t)=s_s(t),
\end{align}
\end{subequations}
is that $E_{1}=E_{2}=....=E_{N}=E_s$. If this condition is satisfied, the synchronous solution $x_s(t)$ obeys,
\begin{subequations}\label{sss}
\begin{align}
\dot{x}_s(t)= & F(x_s(t)) +  \sigma^A (E_s - \varsigma^T x_s(t))  s_{s}(t) \varsigma, \label{sssa} \\
\dot{s}_{s}(t)= & -c_1 s_{s}(t)+c_2 (1-s_{s}(t))[1+\tanh((\varsigma^T x_s(t-\tau)-v_{th})/v_{sl})]. \label{sssb}
\end{align}
\end{subequations}
Note that differently from the case considered in Secs. I, II, and III, the synchronous solution (\ref{xxx}) does not obey the same equation as that of an isolated system.
Our goal in this section is to study stability of the synchronous solution (\ref{xxx}) for the hypernetwork (\ref{IAP}). In order to do that, we linearize the set of equations (\ref{IAP}) about (\ref{xxx}), obtaining
\begin{subequations}\label{LIAP} \begin{align}
\delta \dot{x}_i(t)= & [DF(x_s(t)) -  \sigma^A \Gamma s_s(t)] \delta x_i(t) + \sigma^A  \varsigma (E_s-\varsigma^T x_s(t)) \sum_{j=1}^N A'_{ij} \delta s_{ij}(t)  +  \sigma^B  \sum_{j=1}^N B'_{ij} \Gamma [\delta x_j(t)-\delta x_i(t)],\\
\delta \dot{s}_{ij}(t)= & -c_1 \delta s_{ij}(t)- c_2 S(\varsigma^T x_s(t-\tau)) \delta s_{ij}(t) + c_2 (1-s_s(t)) DS(\varsigma^T x_s(t-\tau)) \varsigma^T \delta x_j(t-\tau),
\end{align} \end{subequations}
where the matrices $A'=\{A'_{ij}\}$ and $B'=\{B'_{ij}\}$ are such that $A'_{ij}={(k^A_i)}^{-1} A_{ij}$ and $B'_{ij}={(k^B_i)}^{-1} B_{ij}$ and we have used the properties that $\sum_j A'_{ij}=1$ and $\sum_j B'_{ij}=1$. We introduce the perturbation $\delta \sigma_i(t)=\sum_j A'_{ij} \delta \sigma_{ij}(t)$, $i=1,...,N$. By multiplying (\ref{LIAP}b) by $A'_{ij}$ and summing over $j$, we can rewrite (\ref{LIAP}) as
\begin{subequations}\label{LIAPR} \begin{align}
\delta \dot{x}_i(t)= & [DF(x_s(t)) -  \sigma^A \Gamma s_s(t)] \delta x_i(t) + \sigma^A  \varsigma (E_s-\varsigma^T x_s(t)) \delta s_{i}(t)  +  \sigma^B  \sum_{j=1}^N B'_{ij} \Gamma [\delta x_j(t)-\delta x_i(t)],\\
\delta \dot{s}_{i}(t)= & -c_1 \delta s_{i}(t)- c_2 S(\varsigma^T x_s(t-\tau)) \delta s_{i}(t) + c_2 (1-s_s(t)) DS(\varsigma^T x_s(t-\tau)) \varsigma^T \sum_j A'_{ij} \delta x_j(t-\tau),
\end{align} \end{subequations}
We can now introduce the $(n+1)$-vectors $\delta \tilde{x}_i(t)=[\delta {x}_i(t)^T, \delta s_i(t)]^T$, $i=1,...,N$ and the $(N(n+1))$-vector $\delta \tilde{x}(t)=[\delta \tilde{x}_1(t)^T, \delta \tilde{x}_2(t)^T,...,\delta \tilde{x}_N(t)^T]^T$. Then,
following Sec. II, we can rewrite the set of equations (\ref{LIAP}) in vectorial form as follows,
\begin{equation}\label{LIAPV}
\delta \dot{\tilde x}(t) =  I_N \otimes [D{\tilde{F}}_1(x_s(t), x_s(t-\tau), s_s(t))] \delta \tilde{x}(t) +    A' \otimes   D{\tilde{F}}_2(x_s(t-\tau),s_s(t))) \delta \tilde{x}(t-\tau)  +  \sigma^B   {L^B}' \otimes \tilde{\Gamma} \delta \tilde{x}(t),
\end{equation}
where the Laplacian matrix ${L^B}'=\{{L^B}'_{ij}\}=\{B'_{ij}-\delta_{ij}\}$ and the $(n+1)$-square matrices
\begin{eqnarray}D{\tilde{F}}_1(x_s(t), x_s(t-\tau), s_s(t))=
{\small\small\small{ \begin{bmatrix} DF(x_s(t)) -  \sigma^A \Gamma s_s(t) &
+ \sigma^A  \varsigma (E_s-\varsigma^T x_s(t))  \cr
0 &  -c_1 - c_2 S(\varsigma^T x_s(t-\tau))  \cr
\end{bmatrix}}},
\end{eqnarray}
\begin{eqnarray}D{\tilde{F}}_2(x_s(t-\tau), s_s(t))=
{\small\small\small{ \begin{bmatrix} 0 &
0  \cr
\varsigma^T c_2 (1-s_s(t)) DS(\varsigma^T x_s(t-\tau)) &  0  \cr
\end{bmatrix}}},
\end{eqnarray}
\begin{eqnarray}\tilde{\Gamma}=
{\small\small\small{ \begin{bmatrix} \Gamma &
0  \cr
0 &  0  \cr
\end{bmatrix}}}.
\end{eqnarray}

As can be seen, the structure of the linearized equations (\ref{LIAPV}) is quite similar to that of Eq. (\ref{lin2}) in Sec.\ II. The main difference with Eq. (\ref{lin2}) is that in the case above, one of the two coupling matrices, namely $A'$, is not a Laplacian matrix, as the entries along each row of the matrix $A'$ sum to one and not to zero. We now wonder  whether the stability problem (\ref{LIAPV}) can be reduced in a low-dimensional form. 
As for the case of Eq. (\ref{lin2}), the main difficulty is that in general it is  impossible to decouple Eq. (\ref{LIAPV}) in $N$ independent blocks. One possibility, which we do not give further consideration in what follows, is that the two matrices $A'$ and ${L^B}'$ commute. Another possibility is that the matrix ${L^B}'$ belongs to class $\mathcal{C}$. If this is the case, then the matrix  ${L^B}'$ can be diagonalized as in Eq. (\ref{WPW}), i.e.,  ${L^B}'=W (I_N^*-I_N) W^{-1}$, where the matrix $W$ is an invertible matrix with the rightmost column being equal to the vector $[1,1,...,1]$ and $I_N^*$ is a diagonal matrix with all the entries on the main diagonal being equal to zero except the one in the rightmost column being equal to one (see Sec. IID). Under these assumptions, the matrix $\Xi=W^{-1} A' W$ has the form
\begin{eqnarray}\Xi= 
{\small\small\small{ \begin{pmatrix} \Xi_{11} &
\Xi_{12}  &  \cdots & \Xi_{1(N-1)} & 0 \cr
\Xi_{21} &\Xi_{22} & \cdots & \Xi_{2(N-1)} & 0   \cr
& & \vdots &  \cr
\Xi_{(N-1)1} & \Xi_{(N-1)2} & \cdots & \Xi_{(N-1)(N-1)}  & 0\cr
\Xi_{N1} & \Xi_{N2} & \cdots & \Xi_{N(N-1)} & 1
 \end{pmatrix}}},
\end{eqnarray}
 from which we see that similarly to Sec. IIC, the linearized problem (\ref{LIAPV}) can be decoupled into a drive subsystem and a response subsystem, with the response subsystem corresponding to perturbations tangent to the synchronization manifold (\ref{xxx}) and the drive subsystem corresponding to perturbations transverse to the synchronization manifold.

It is known from the literature that in the visual cortex \cite{Fekuda2} and in the posterior part of the putamen \cite{Fekuda1}, small groups of neurons are likely to form dense and uniform clusters of gap-junctions. 
Hence, assuming that the 
network ${L^B}'$ is of class $\mathcal{C}$ can be appropriate to model such agglomerates of neurons or small subsets of them.
 Therefore, as an example,
 we consider a small group of $N$ neurons connected by a dense  ${L^B}'$ network of  gap junctions, with ${L^B}' \in \mathcal{C}$.
Under these assumptions, by diagonalizing the $(N-1)$-dimensional subspace of transverse perturbations (see Sec. II), the high-dimensional problem (\ref{LIAP}) can be reduced into the low-dimensional form,
\begin{equation}\label{LIAPLD}
 \dot{\vartheta}(t) = [D{\tilde{F}}_1(x_s(t), x_s(t-\tau), s_s(t)) -  \sigma^B    \tilde{\Gamma}] \vartheta(t)
 +  {\lambda^A}'_k   D{\tilde{F}}_2(x_s(t-\tau),s_s(t))) \vartheta(t-\tau),
 \end{equation}
$k=1,2,...,(N-1)$, where 
$\{{\lambda^A}'_k\}$, $k=1,...,N$, are the eigenvalues of the matrix $A'$.  By construction, the matrix $A'$ has one eigenvalue,  ${\lambda^A}'_N=1$, with associated eigenvector $[1,1,...,1]$. This eigenvector represents perturbations that are tangent to the synchronous solution, hence it is not relevant in determining stability of the synchronous solution (\ref{xxx}).

In the more general case in which ${L^B}'$ does not belong to class $\mathcal{C}$ and the two matrices ${L^B}'$ and $A'$ do not commute, stability of the synchronous solution results in a much more complex problem, for which (\ref{LIAP}) cannot be reduced in a low dimensional form and we expect a higher degree of complexity. 
The study of this case is beyond the scope of this paper.



We run numerical simulations in which each individual system is described by the FitzHugh-Nagumo model, $n=2$,
\begin{equation}
F({{x}})=\left[\begin{array}{c}
    10[x_{1}(x_1-0.1)(1-x_1)-x_2+0.2]  \\
    x_{1} - 0.5 x_{2} \\
  \end{array} \right] \label{FHN}
\end{equation}
and we set $v_{th}=0.3$, $v_{sl}=10^{-2}$, $c_1=c_2=10$, $E_s=1$, $\sigma^A=1$, $\tau=1$. In Fig.\ 6(a) we plot the time evolution of the synchronous evolution, obtained by integrating Eq.\ (\ref{sss}) for this particular choice of the function $F$ in (\ref{FHN}) and of the parameters. We further set $\sigma^B=0.9$ and calculate the maximum Lyapunov exponent associated with the low-dimensional system (\ref{LIAPLD}) as a function of the parameter ${\lambda^A}'$. 
This corresponds to a master stability function (MSF), which is plotted in Fig.\ 6(b) for the case that its argument is real.
As can be seen from Fig.\ 6(b), the MSF curve crosses the $0$-ordinate line at two distinct values of the abscissa, which we found to be approximately equal to $-0.74$ and $1.1$ (in the figure, the $0$-ordinate and the $1$-abscissa lines are plotted as dashed lines). Thus a necessary and sufficient condition for stability of the synchronous solution is that $-0.74 \leq {\lambda^A}'_i \leq 1.1$, $i=1,...,(N-1)$. If we assume $A'_{ij} \geq 0$,  we have by the  Perron-Frobenius theorem that $|{\lambda^A}'_i| \leq 1$, $i=1,...,N$, where $1$ is the Perron-Frobenius eigenvalue of the matrix $A'$, and the necessary and sufficient condition for stability reduces to $-0.74 \leq {\lambda^A}'_{min}$, where ${\lambda^A}'_{min}= \min_{i=1,...,(N-1)} {\lambda^A}'_i$.

We finally run simulations of the full nonlinear hypernetwork described by Eqs. (\ref{IAP},\ref{sij},\ref{FHN}). We set the initial conditions for $x^1_i$ and $x^2_i$, $i=1,...,N$ and for $s_{ij}$, $i,j=1,...,N$ to be random numbers drawn from a uniform distribution in the range $(0, 0.2)$. We consider that the network of chemical synapses is the network of $N=6$ nodes and  $9$ directed links represented in gray in Fig. 2, i.e.,  the entries of the matrix $A$ are $A_{ij}=1$ if there is a  gray direct arrow from node $j$ to node $i$ in the figure, $A_{ij}=0$ otherwise. The spectrum of the corresponding matrix $A'$ is real and ${\lambda^A}'_{min}=-\sqrt{2}/2>-0.74$. We set the network of chemical synapses to be such that $B_{ij}=b_j=j$, $i,j=1,...,6$ (note that the particular choice of the values of $b_j$, $j=1,...,N$ affects neither the spectrum of the matrix ${L^B}'$ nor the low-dimensional equation (\ref{LIAPLD})). We evolve the hypernetwork (\ref{IAP},\ref{sij},\ref{FHN}) from $t=0$ to $t=500$. We monitor the quantity $E(t)$, defined in Eq. (\ref{error}). As expected, we observe that after a transient, $E(t) \rightarrow 0$. We repeat the same experiment for the case that $A_{ij}=1$ if $|i-j|=1$ and $A_{ij}=0$ otherwise. For this case, the spectrum of the corresponding matrix $A'$ is real but ${\lambda^A}'_{min}=-1<-0.74$, thus predicting that the synchronous solution is unstable. This is confirmed by our numerical experiments, showing that, when the full nonlinear system (\ref{IAP},\ref{sij},\ref{FHN}) is integrated from initial conditions that are close to the synchronous state (\ref{sss}), $E(t)$ does not converge to $0$.

\begin{figure}
\centering
\includegraphics[width=3.5in]{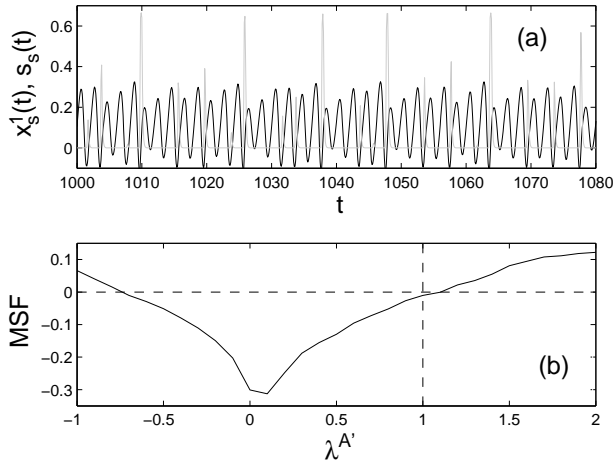}
\caption{(a) Time evolution of the synchronized solution for the system (\ref{IAP}), obtained by numerically integrating Eqs. (\ref{sss},\ref{FHN}) with $v_{th}=0.3$, $v_{sl}=10^{-2}$, $c_1=c_2=10$, $E_1=E_2=...=E_N=E_s=1$, $\sigma^A=1$, $\tau=1$.  $x^1_s(t)$ is plotted in black and $s_s(t)$ is plotted in gray. (b) Plot of the master stability function corresponding to the low-dimensional system (\ref{LIAPLD}) versus the parameter ${\lambda^A}'$ for the case that ${\lambda^A}'$ is real. The parameters are the same as in (a), $\sigma^B=0.9$.   }
\label{Grafo}
\end{figure}



\section{Conclusion and discussion}

In this paper we have studied synchronization of coupled dynamical systems when different types of interactions are simultaneously present.
Our study applies to any situation where the individual units interact through different coupling mechanisms.  For example, neurons in the brain are known to  be connected through both  electrical gap junctions and chemical synapses,  \cite{IZH,Ad:Pr:Dh,SPCENN,MORCGCENS}. Also, our study encompasses a situation where different coupling functions correspond to different interaction delays.

In our formulation, a set of identical dynamical systems are coupled through the connections of two or more distinct
networks (each of which corresponds to a distinct coupling function) and we refer to such a system as a dynamical hypernetwork.  
We first focus on the case of a hypernetwork formed of $m=2$ networks and we seek to obtain necessary and sufficient conditions for synchronization. In Sec. II we try to reduce the stability problem in a master stability function form. Though a solution in this form seems to be not available in general, we show that such a reduction is possible in three cases of interest: (i) the Laplacian matrices associated with the two networks commute; (ii) one of the two networks is unweighted and fully connected; (iii) one of the two networks is such that the coupling strength from node $j$ to node $i$ is a function of $j$ but not of $i$, with case (ii) being a subcase of (iii). We  introduce a unique master stability function that determines stability for all three cases. 
Also, we define the class $\mathcal{C}$ of networks for which the reduction is always possible, independent of the structure of the other network.

We note that in many situations, such as, e.g., in biological networks, different types of interactions are typically present, but the couplings may vary in time due to changing environmental conditions, making satisfaction of either one of conditions (i), (ii), or (iii) difficult. On the one hand, this highlights a limitation of the master stability function approach that does not seem to be applicable to situations of arbitrary complexity (see also e.g., \cite{NSG}). On the other hand, it poses the fascinating challenge of defining alternative tools to addressing stability for the case of arbitrary hypernetworks.
We also point out here that we cannot exclude the existence of other conditions to be satisfied simultaneously by both matrices $A$ and $B$ (e.g., for either the hypernetwork  (\ref{AB}) or (\ref{IAP})) that allow a reduction of the stability problem in a low-dimensional form.

In Sec. IV we have proposed a generalization of our stability results to hypernetworks formed of $m$ networks. In Sec. V we have shown the possibility of generalizing our approach to hypernetworks of coupled systems, which cannot be cast into the specific form of Eqs.\ (\ref{AB}). As an example of interest, we have studied synchronization of a neural hypernetworks for which the connections can be either chemical synapses or electrical gap junctions.  The results of this paper could be also easily extended to study synchronization of dynamical hypernetworks of coupled discrete-time systems. 


\section*{Appendix: The special case of hypernetworks of $N=2$ nodes}

{In this appendix we show that for hypernetworks of $N=2$ nodes, the stability problem can always be reduced in a low-dimensional form. We start by considering that $N$ is an arbitrary number and that the hypernetwork is formed of $m=2$ networks (Eq. (\ref{AB})). The generalization to the case of $m>2$ networks is straightforward.
}

{We look at Eq. (\ref{LALB}). In general, a case of interest is that one of the two Laplacian matrices, say $L^A$, can be rewritten as,
\begin{equation}\label{dec}
L^A=k_1 L^{A1}+k_2 L^{A2},
\end{equation}
where the matrix $L^{A1}$ belongs to $\mathcal{C}$ (i.e., it is in the form of the matrix (\ref{LA})) and the matrix $L^{A2}$ commute with $L^{B}$, that is $L^{A2}= V \Lambda^A V^{-1}$ and $L^B= V \Lambda^B V^{-1}$, where $\Lambda^A$ and $\Lambda^B$ are diagonal matrices. Under the condition (\ref{dec}), Eq. (\ref{lin2}) can be rewritten as,
\begin{equation} \label{cas2}
\begin{split}
\delta \dot{x}(t) = & I_N \otimes DF(x_s(t)) \delta {x}(t)  +   \sigma^A k_1 L^{A1} \otimes DG(x_s(t-\tau_g)) \delta {x}(t-\tau_g) \\
+ & \sigma^A k_2 L^{A2} \otimes DG(x_s(t-\tau_g)) \delta {x}(t-\tau_g) + \sigma^B {L^B} \otimes DH(x_s(t-\tau_h)) \delta {x}(t-\tau_h).
\end{split}
\end{equation}
Following Sec. IIC, it can be shown that a necessary and sufficient condition for stability of the synchronous solution for the hypernetwork (\ref{cas2}) is that the maximum Lyapunov exponent of the low-dimensional equation,
\begin{equation} \label{case2ld}
\dot{\theta}_k(t) =  DF(x_s(t)) \theta_k(t) + \sigma^A (k_2 \lambda^A_k-k_1 \bar{a})  DG(x_s(t-\tau_g)) \theta_k(t-\tau_g) +\sigma^B \lambda^B_k DH(x_s(t-\tau_h))  {\theta}_k(t-\tau_h),
\end{equation}
is negative for $k=1,...,(N-1)$, where $\bar{a}=\sum_{j=1}^N a_j$, $\lambda^A_k$ and $\lambda^B_k$ are respectively the (complex) eigenvalues of the matrices $L^A$ and $L^B$ that are associated with the same eigenvectors, i.e., such that $L^A v_k= \lambda^A_k v_k$ and $L^B v_k= \lambda^B_k v_k$. Note that the one eigenvalue $\lambda^A_N=\lambda^B_N=0$ is not relevant in determining stability.
}
{Now the question arises how likely it is that an arbitrary Laplacian matrix $L^A$ can be decomposed in the form (\ref{dec}). In general terms, an $N$-squared matrix is determined by its $N^2$ entries. At the same time, we are allowed $2N$ degrees of freedom in the decomposition (\ref{dec}), i.e., $N$ degrees of freedom in choosing the entries $a_1,a_2,...,a_N$ of the $\mathcal{C}$-matrix $L^{A1}$ and $N$ degrees of freedom in choosing the eigenvalues of the matrix $L^{A2}$. It follows that only in the case that $N=2$, a decomposition in the form (\ref{dec}) is guaranteed irrespective of the choice of the two Laplacian matrices $L^A$ and $L^B$. This leads to the conclusion that the stability of the synchronous solution for an arbitrary $N=2$-hypernetwork can always be associated with the MLE of the low-dimensional equation (\ref{case2ld}) for $k=1$.
}

\end{document}